**Comment on "Statistical nature of multifragmentation":** In a recent paper, Botvina and Gross [1] sharply criticize the finding by Tõke et al. [2] of the devastating effect of statistical fluctuations on the interpretation of nuclear pseudo-Arrhenius plots. Such plots have been discussed in a series of recent papers [4]. We wish to point out that this criticism has no relevance for the conclusions of Ref. [2] and that, hence, the latter still stand. At the same time, we show that the conclusions of Ref. [1] are unfounded, based on erroneous premises.

The criticism of Ref. [2] by Botvina et al. [1] is based on the following "observations":
(i) Ref. [2] incorrectly refers to the Copenhagen model SMM [3] as the "Berlin SMM model".
(ii) Ref. [2] demonstrates a strong effect of statistical fluctuations based on simulation calculations which utilize the equilibrium-statistical model code GEMINI [5].
(iii) Ref. [2] assumes a binomial fragment distribution even for very small excitation energies.
(iv) Ref. [2] disregards the effects of imperfections of the experimental setup (experimental filter).

The irrelevance of (i) is rather obvious and Ref. [1] offers no arguments to the contrary.

The irrelevance of (ii) is clear from the fact that the use of the Copenhagen model SMM [3] in simulation calculations [2] results in even a stronger effect than that obtained with GEMINI [5]. The authors of Ref. [1] have apparently overlooked this fact, as they do not dispute the applicability of the SMM model.

The irrelevance of (iii) is clear from the fact that singularities in theoretical pseudo-Arrhenius plots are expected [2] to occur at the rather high excitation energies of over 1 GeV. It is also more than debatable whether the approximately 400 MeV of excitation energy, where the calculations in Ref. [2] begin, qualify as a "very small excitation when this *(the fragment production)* is forbidden by energy conservation".

The irrelevance of (iv) is less obvious and, hence, requires a more detailed analysis. Observation (iv) implies that, for excitation energies of 1 MeV/u or so, the MSU Miniball has such a low efficiency that it measures only approximately 33% of the true transverse energy $E_t$ (see Fig. 1a, solid and dot-dashed lines). This unexpectedly strong filtering effect can be traced back (see Fig. 1 in Ref. [1]) to an unusually large $E_t$ predicted by the Berlin MMMC. [6] Ref. [1] attributes the excess transverse energy to "a stronger transversality in MMMC because small relative angles become suppressed in a simultaneous fragmentation". In fact, suppression of small relative angles has no effect on the transversality, as measured by the average value of the $sin^2(\Theta)$ of the emission angle $\Theta$. Indeed, MMMC does predict $<sin^2(\Theta)>\approx 2/3$, as is proper for isotropic emission. It appears, however, that the excess $E_t$ in Fig. 1 of Ref. [1] is due to an improper inclusion of fission fragments in the definition of $E_t$, contrary to the experimental definition. Consequently, the effect of the filter, which Botvina and Gross claim [1] to have discovered, is merely that of the insensitivity of the detectors to fission fragments. These fragments, however, should not have been included in $E_t$ at all.

The conclusion of Ref. [2], now supported by Fig. 3 in Ref. [1], is that the statistical fluctuations inherent in the relationship between the excitation energy $E^*$ and $E_t$ would have destroyed any information about the binomial parameters $p$ and $m$, had the IMF multiplicity distributions been originally binomial. This conclusion builds on the principle that information once lost cannot be retrieved, no matter how poor an experimental instrument is used in the attempt. In contrast, Ref. [1] builds on a proposition that information lost due to statistical fluctuations can be recovered when an imperfect instrument is employed. However, Ref. [1] presents no proof for the general validity of such a radical proposition. To our knowledge, such retrieval of lost information has never been demonstrated.

Lastly, since Ref. [1] repeatedly stresses the superiority of "a proper microcanonical" treatment, we wish to point out that the MMMC [6] itself does not belong to this class of proper microcanonical models. Quite obviously, MMMC does not treat evaporated neutrons microcanonically, which results in a largely underestimated role of neutron channels - in reality the dominant decay modes. Consequently, MMMC incorrectly describes the character of the average correlation between $E_t$ and $E^*$ as an approximate proportionality, whereas the actual, experimentally observed relationship is that of an approximately linear function with a significant offset in $E^*$. For this reason, MMMC is not suitable to describe situations where the dependence of $E_t$ on $E^*$ is essential, which include the case of nuclear pseudo-Arrhenius plots. In contrast, both SMM [3] and GEMINI [5] correctly account for the rather large magnitude of this offset (approx. 340 MeV for the Xe+Au system).

In summary, the conclusion of Ref. [1] regarding a statistical nature of multifragmentation are unfounded. However, it is also our own expectation that nuclear multifragmentation would show statistical features, regardless of whether it is driven thermally or dynamically.

This work has been supported by the U.S. Department of Energy Grant. No. DE-FG02-88ER-40414.

J.Tõke and W.U. Schröder
*Department of Chemistry and Nuclear Science Research Laboratory, University of Rochester, Rochester, New York 14627*